\documentstyle[aps,epsfig,multicol]{revtex}
\begin{document}
\draft
\title{
Interaction of nonuniform elastic waves with two-dimensional electrons
in $Al Ga As - Ga As - Al Ga As$ heterostructures}
\author{D.V.Fil}
\address{Institute for Single Crystals National Academy of Sciences of
Ukraine,
Lenin av. 60 Kharkov 310001 Ukraine\\ e-mail: fil@isc.kharkov.com}
\date{February 10,1999}
\maketitle
\begin{abstract}
An interaction of a double layer electron system realized in an
$Al Ga As - Ga As - Al Ga As$ heterostructure with nonuniform
elastic modes localized in the $Ga As$ layer is considered.
The dependence of the coupling constant on the ratio between the
thickness of the $Ga As$ layer and the wavelength is calculated
for the wave vector directed along the [110] axis, the displacement vector
lying in the (1${\bar 1}$0) plane and the interface boundaries parallel to
the (001) plane. It is shown, that the coupling constant reaches the
maximal value at the wavelength which is of order of the thickness of the
$Ga As$ layer. The renormalization of the velocity of the elastic modes
is found for the
case, when the electron system is in the fractional quantum Hall regime.
It is shown, that for certain modes the dependence of the velocity
shift on the wave vector is modified qualitatively
under a transition of the electron system to the state, which corresponds to
the Halperin wave function.
\end{abstract}
\begin{multicols}{2}
Recently the surface acoustic waves (SAW)  have been
widely used for an investigation of the dynamical properties
of two-dimensional electron layers in heterostructures
$Al Ga As$ \cite{1,2}. Since the $Al Ga As$ is a piezoelectric
material, the SAW generates an alternating electric field.
The interaction of that field with the two-dimensional electrons results
in a renormalization of the velocity and the absorption of the SAW.
The frequency and the momentum dependence of the conductivity of
the electron system can be obtained from the measurements of
the velocity shift and the absorption coefficient.
In particular, the power of the method has been  demonstrated
within the study of the quantum Hall systems for which the conductivity depends
considerably on the external magnetic field.

Since the electron layer is placed at a certain distance $d_0$
from the surface of the sample, the coupling of the SAW with electrons
depends on the parameter $q d_0$  (where $q$ is the wave vector of the
elastic mode). An exponential reduction of the matrix elements of the
coupling takes place if the parameter $q d_0$ increases. Therefore, the
SAW method
is not applicable  for a study of the dynamical properties of the electron
system at large $q$. In that case a nonuniform elastic wave, which is
localized near the electron layer due to the acoustic nonhomogeneity of
the heterostructure, can be used.

To specify the system, in which the elastic wave may be localized, we consider
the  heterostructure   $Al_x Ga_{1-x} As - Ga As - Al_x Ga_{1-x} As$
which incorporates two electron layers situated at the
$Al Ga As - Ga As$ interfaces. Similar heterostructures have been used for
the investigation of the quantum Hall effect in the double-layer electron
systems \cite{3,4}.

Several types of the elastic modes localized in a central layer
exist in the structure considered.
The interaction of the two-dimensional electrons with the nonuniform
transverse elastic wave with the displacement vector parallel to the
interface boundaries
has been studied in \cite{5}.
In this paper we continue the investigation of Ref. \cite{5} and consider
the nonuniform elastic waves for which the displacement vector is in the
plane determined by the wave vector and the normal direction to the
interface boundaries (these waves as well as the SAW are
polarized elliptically). We found that the interaction of such waves with
electrons may be much stronger then of the waves considered in
Ref. \cite{5}.

\section{Geometry of the model and the dispersion equations}

Let us consider the system for which the $GaAs$ layer with the
thickness $2 a$ is situated between two $Al_{0.3} Ga_{0.7} As$
layers  with the thicknesses much larger than the
wavelength of the elastic mode. We specify the case of the interface
boundaries  parallel to the (001) plane and the wave vector of the
elastic mode directed along the [110] axis. We will use the fact
that the elastic moduli are practically independent on the
concentration  of $Al$ (we set them the same for the both
media). The acoustical nonuniformity of the system is caused by the
difference of the densities
($\rho_1 $=5.3 g/cm$^3$ for $Ga As$,
$\rho_2$=4.8 g/cm$^3$ for $Al_{0.3} Ga_{0.7} As$).

We consider the elastic mode with the displacement vector
components
\begin{equation}
u_i({\bf r},z,t)=u_i(z) e^{i {\bf q r} - i \omega  _q t},
\label{1}
\end{equation}
where $\omega_q =v q$, $v$ is the elastic mode velocity,
$i=x,z$, the $x$ axis is chosen along the
[110] direction, the $z$ axis, along the [001] direction,
the ${\bf r}$ vector is in the (001) plane. The wave equations for
the $u_i(z)$  components have the form
\begin{eqnarray}
(c_{44}\partial^2_z - c^{'}_{11} q^2 + \rho_{\alpha} \omega^2) u_x +
i q (c_{12}+c_{44})\partial_z u_z =0\cr
i q (c_{12}+c_{44})\partial_z u_x+
(c_{11}\partial^2_z - c_{44} q^2 + \rho_{\alpha} \omega^2) u_z =0 ,
\label{2}
\end{eqnarray}
where $c^{'}_{11}=0.5(c_{11}+c_{12}+2 c_{44})$, $\alpha$=1,2 corresponds to
the medium number. The solution of Eqs.(\ref{2}) has the form
\begin{equation}
u_i^\alpha(z) = \sum_k A_{i k}^\alpha \exp y_k^\alpha z q,
\label{3}
\end{equation}
where $y_k^\alpha$ are the roots of the equation
\begin{equation}
y^4 + 2 b_\alpha y^2 +c_\alpha =0
\label{4}
\end{equation}
with
\begin{eqnarray}
b_\alpha ={1\over 2 c_{11} c_{44}}
[(c_{12}+c_{44})^2 + c_{11}(\rho_\alpha v^2 -c^{'}_{11}) \cr +
c_{44}(\rho_\alpha v^2 -c_{44})]  \cr
c_\alpha ={1\over  c_{11} c_{44}}
(\rho_\alpha v^2 -c^{'}_{11})
(\rho_\alpha v^2 -c_{44}).
\label{5}
\end{eqnarray}

If we take into account the equivalence of the elastic
moduli for two media,
the boundary conditions are reduced to the requirement
of the continuity of $u_i$ and $\partial_z u_i$
at the interfaces
The localized mode corresponds to the solution for which
the displacement
approaches to zero at $z\to\pm\infty$.
The structure of the localized solution in the $AlGaAs$ layers is
similar to the
structure of the SAW on the surface of the cubic crystal
(see, for instance, Ref. \cite{6}). Such a solution arrears if
Eq. (\ref{4}) at $\alpha =2$ does not have imaginary roots.
If the elastic moduli satisfy the inequality
\begin{equation}
(c_{12}+c_{44})^2 - c_{11} (c^{'}_{11}-c_{44})<0 ,
\label{6}
\end{equation}
the localized solution emerges at
$v<c_{44}/\rho_2$. If the opposite inequality is satisfied,
the velocity of the localized mode
$v<v_{m2}$,
where $v_{m2}$ is the root of the equation
$D_2(v)=b_2^2-c_2=0$
($v_{m2}<c_{44}/\rho _2$).
For the system considered (the values of the elastic moduli are given
below) the second case is realized.
At $v<v_{m2}$
Eq. (\ref{4}) for $\alpha =2$ gives
\begin{equation}
y=\pm(\lambda \pm i\varphi ),
\label{7}
\end{equation}
where
\begin{eqnarray}
\lambda =\sqrt{(\sqrt{c_2}-b_2)/2}\cr
\varphi  =\sqrt{(\sqrt{c_2}+b_2)/2} .
\label{8}
\end{eqnarray}
For the medium 1 the solution of Eq. (\ref{2})
which satisfy the boundary conditions corresponds to the cases, for which
Eq. (\ref{4}) has two real and two imaginary roots or four imaginary roots
\begin{eqnarray}
y_{1,2} =\pm\kappa =\pm\sqrt{\sqrt{D_1}-b_1} \cr
y_{3,4} =\pm i\xi =\pm i\sqrt{\sqrt{D_1}+b_1} ,
\label{9}
\end{eqnarray}
where $D_1=b_1^2-c_1$. (The $\kappa$ is a real value at
$v>c_{44}/\rho_1$, and an imaginary value at
$v_{m1}<v<c_{44}/\rho_1$;  $v_{m1}$ is the root of equation
$D_1(v)=0$)

There are two types of the solutions of the wave equations, which satisfy
the boundary conditions
(we will refer on them as I and II)
\begin{eqnarray}
u_x^p(z)=C^p f_x^p(z)\cr
  u_z^p(z)=i C^p f_z^p(z) ,
\label{10}
\end{eqnarray}
where $C^p$ is the normalization factor, $p$=I,II.
In Eq.(\ref{10})
$f_x^{\rm{I}}(z)$ is the odd function and
$f_z^{\rm{I}}(z)$ is the even function, while
$f_x^{\rm{II}}(z)$ is the even function and
$f_z^{\rm{II}}(z)$ is the odd function.

The dispersion equation for $v$ is obtained from the common linear problem
on the factors $A_{ik}^\alpha $.
For the mode I the dispersion equation reads as
\begin{equation}
R_1 {\tanh} \kappa q a \ {\tan} \xi q a +
R_2 {\tanh} \kappa q a  +
R_3  {\tan} \xi q a +
R_4 =0 .
\label{11}
\end{equation}
For the mode II it is modified to
\begin{equation}
R_1 {\coth} \kappa q a \ {\cot} \xi q a -
R_2 {\coth} \kappa q a  +
R_3  {\cot} \xi q a -
R_4 =0  ,
\label{12}
\end{equation}
where
\begin{eqnarray}
\nonumber
R_1=(\kappa m_\kappa +\xi m_\xi)
(\varphi m_\lambda -\lambda m_\varphi)\cr
R_2= m_\varphi m_\xi (\lambda ^2 +\varphi ^2) -
\varphi \xi (m_\varphi ^2 +m_\lambda^2) \cr -
\kappa m_\kappa (\xi m_\varphi -\varphi m_\xi) \cr
R_3= -m_\varphi m_\kappa  (\lambda ^2 +\varphi ^2) +
\varphi \kappa  (m_\varphi ^2 +m_\lambda^2) \cr +
\xi m_\xi (\varphi  m_\kappa  -\kappa  m_\varphi )\cr
R_4=(\kappa m_\xi - \xi m_\kappa )
(\varphi m_\lambda +\lambda m_\varphi)\cr
m_\kappa={\kappa (c_{12}+c_{44})\over
c_{11}\kappa ^2 - c_{44} + \rho_1 v^2}\cr
m_\xi={\xi (c_{12}+c_{44})\over
-c_{11}\xi ^2 - c_{44} + \rho_1 v^2}  \cr
m_\lambda ={\lambda c_{44} (R-1)\over c_{12}+c_{44}} \cr
m_\varphi =-{\varphi  c_{44} (R+1)\over c_{12}+c_{44}} \cr
R=\sqrt{c_{11}(\rho_2 v^2 -c^{'}_{11})\over c_{44} (\rho_2 v^2- c_{44})}.
\label{}
\end{eqnarray}

\begin{figure}
\narrowtext
\centerline{\epsfig{figure=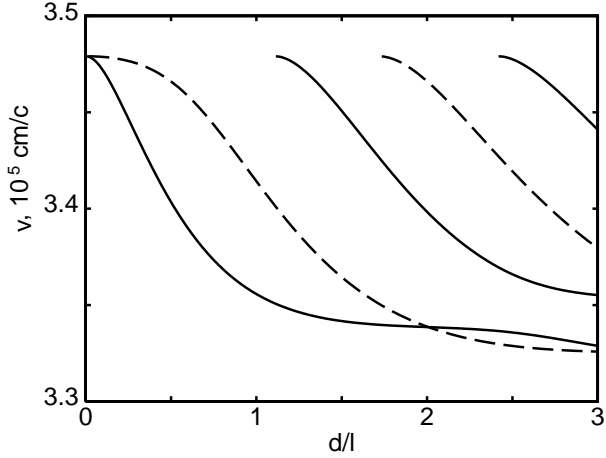,width=8cm}}
\vspace{0.5cm}
\caption{Dependences of the velocities of the elastic modes on the
ratio between the thickness of the central layer $d$ and
the wavelength $l$. Solid lines indicates the type I modes;
dashed lines, the type II modes.}
\label{fig1}
\end{figure}

\begin{figure}
\narrowtext
\centerline{\epsfig{figure=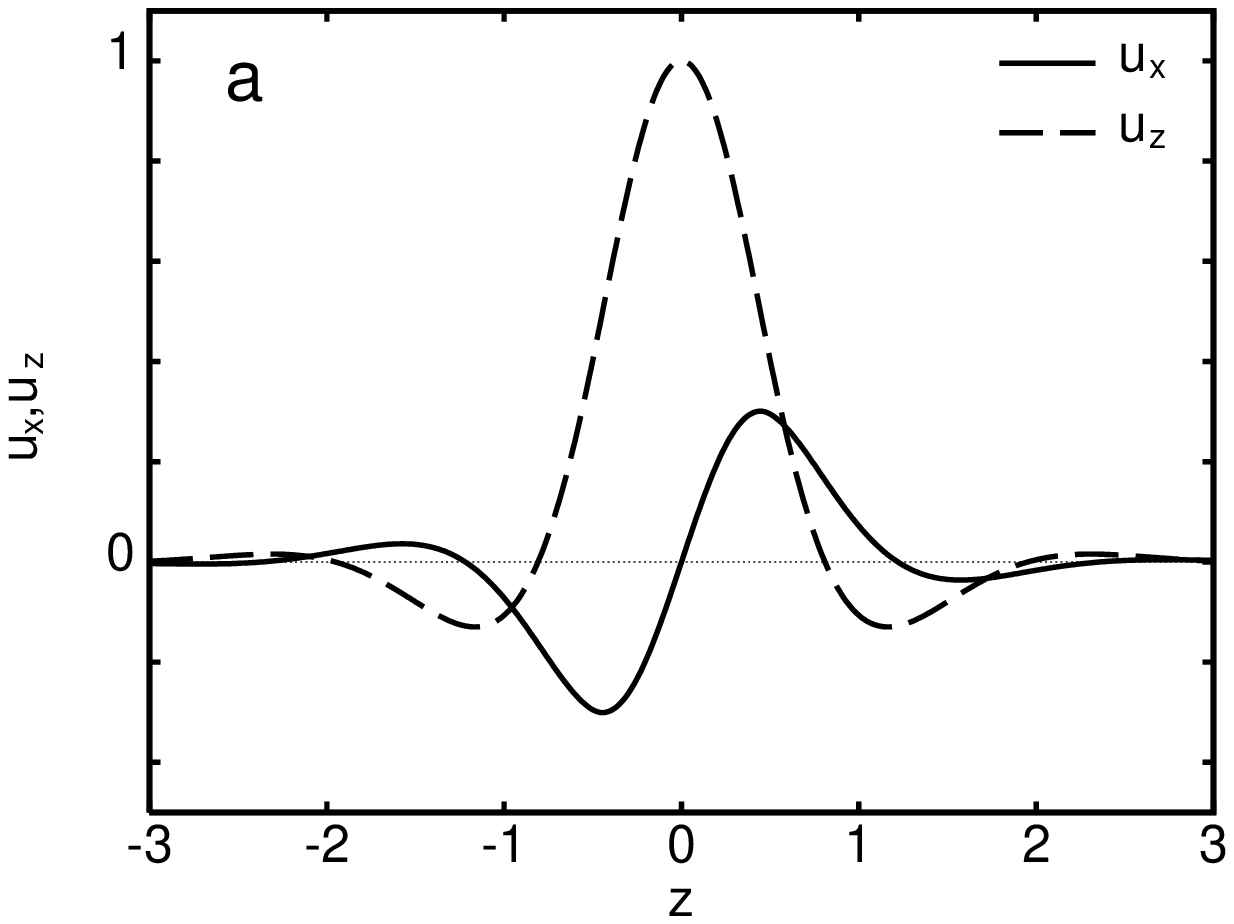,width=8cm}}
\vspace{0.5cm}
\centerline{\epsfig{figure=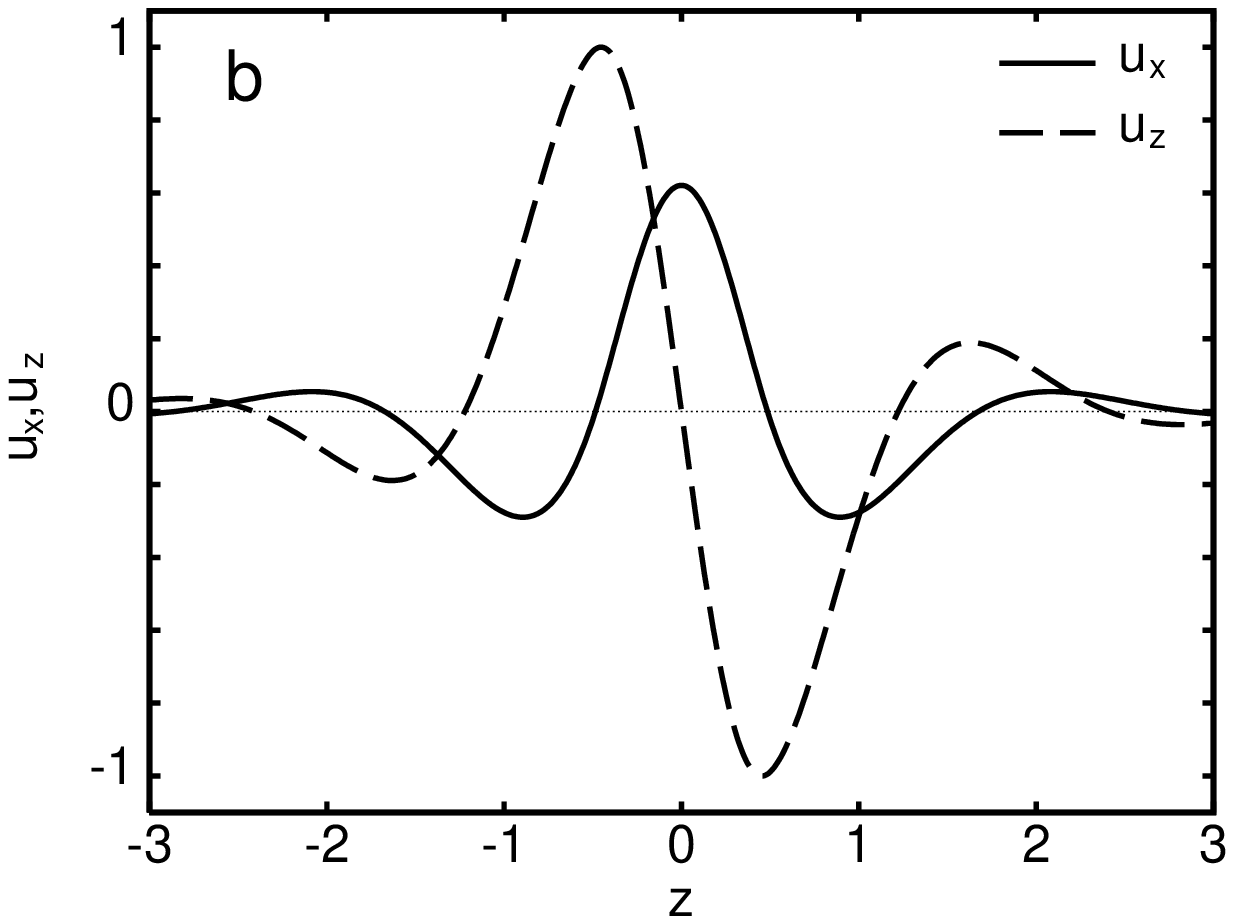,width=8cm}}
\vspace{0.5cm}
\caption{
Dependences of the  displacement vector components
(in relative units)
on $z$ (in units of the wavelength) at $d/l=1$;
a, the type I mode; b, the type II mode.}
\label{fig2}
\end{figure}

Numerical solutions of Eqs. (\ref{11},\ref{12}) for the parameters
$c_{11}=12.3$, $c_{12}=5.7$, $c_{44}=6.0$ (all in 10$^{11}$ dyn/cm$^2$)
versus the ratio between the thickness of the central layer
($d=2 a$) and the wavelength are shown in Fig.~\ref{fig1}.
One can see from the dependences presented, there are two gapless
modes (one of them is of the type I, while the other is of the type II)
in the system. When the wavelength becomes shorter, additional modes
with higher frequencies emerge.
The interaction with the gapless modes
is only considered below.
The quantities  $f_i^p(z)$ can be written through elementary functions
of $z$,
but the expressions contain a complicate dependence on
$q$ and $v$. Therefore we will not present here the analytical
expressions for $f_i^p(z)$. At an example, the plots
of  $u_x(z)$, $u_z(z)$ at $d/l=1$ are shown in Fig.~\ref{fig2} .

\section{Piezoelectric coupling constants}

Let us calculate the renormalization the velocity of the nonuniform
elastic mode, caused by the interaction with the double-layer electron
system with the coordinates of the electron layers
$z_{1(2)}=\pm a$.
We use the approach similar to Ref.
\cite{7}.
Let us write the Hamiltonian for the elastic waves in terms of the
phonon creation and annihilation operators
($b^+$, $b$)
\begin{eqnarray}
H_{\rm u}=\sum_{\bf q} \omega _q (b^+_{\bf q} b_{\bf q} +{1\over 2}),
\label{13}
\end{eqnarray}
The Hamiltonian of the electron-phonon interaction is chosen in the
form
\begin{eqnarray}
H_{\rm int} = {1\over  \sqrt{S}} \sum_{{\bf q},m} \int d^2 r
\ g_{q m}\Psi^+_{{\bf r} m} \Psi_{{\bf r} m} e^{i {\bf q r}}
(b_{\bf q} + b^+_{-{\bf q}}),
\label{14}
\end{eqnarray}
where $\Psi^+(\Psi)$ are the electron creation (annihilation) operators,
$m$, the number of the electron layer,
$S$, the area of the layer.

To obtain the matrix elements $g_{q m}$ we write the interaction
of the elastic wave with the electrons in the form
\begin{eqnarray}
H = \sum_{m} \int d^2 r \ e \varphi_{{\bf r} m} \Psi^+_{{\bf r} m}
 \Psi_{{\bf r} m},
\label{15}
\end{eqnarray}
where $\varphi_{{\bf r} m}$ is the scalar potential of the electric field,
generated by the elastic wave in the layer $m$.
The value of $\varphi $
is determined by the solution of the Poisson equation
\begin{eqnarray}
\Delta \varphi = -(4\pi/ \epsilon)
\beta_{i,j k} \partial_i u_{j k} ,
\label{16}
\end{eqnarray}
where $\epsilon $ is the dielectric constant,
$\beta_{i,j k}$ is the piezoelectric tensor,
$u_{j k}$ is the strain tensor.
Under the choice of the $x$, $y$ and $z$ axes along
the [100], [010] and [001] directions correspondingly the
$\hat{\beta}$ tensor has nonzero (and all equal to the same value $\beta$)
components for
$i\ne j\ne k$.
Under substitution of Eq.(\ref{10}) into Eq.(\ref{16})
we obtain the following equation for the Fourier-component
of the electric potential
\begin{equation}
(\partial_z^2 - q^2) \varphi_q(z)= i(4\pi\beta/\epsilon) C^p g^p(z),
\label{17}
\end{equation}
where
\begin{equation}
g^p(z)= q^2 f^p_z(z)- 2  q \partial_z f^p_x(z).
\label{18}
\end{equation}

\begin{figure}
\narrowtext
\centerline{\epsfig{figure=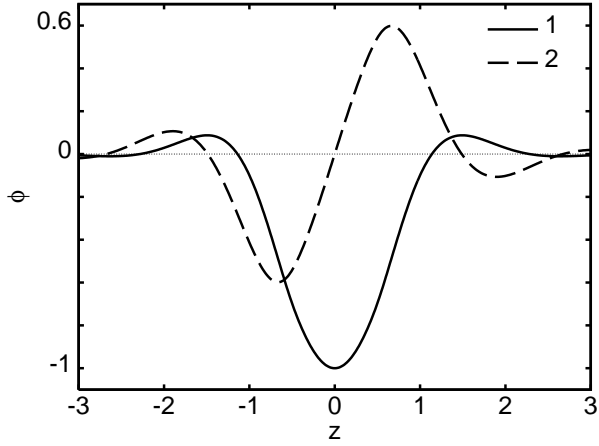,width=8cm}}
\vspace{0.5cm}
\caption{
Dependence of the  Fourier component of the electric potential
(in relative units)
on $z$ (in units of the wavelength) at $d/l=1$;
1, the type I mode; 2, the type II mode.}
\label{fig3}
\end{figure}

We assume that the
$\epsilon $ and $\beta $ parameters are same for the whole system.
Then the boundary conditions on $\varphi$ reduce to the requirement of
continuity of  $\varphi_q(z)$ and
 $\partial_z \varphi_q (z)$ at the interfaces.
The solution of Eq.
(\ref{17})  with the boundary conditions  has the form
\begin{equation}
\varphi_q(z) = i (4\pi \beta /\epsilon ) C^p \chi_p(z),
\label{19}
\end{equation}
where $\chi_{\rm I}(z)$ is an even function and
$\chi_{\rm II}(z)$ is an odd function.
We do not present the analytical expressions for the
$\chi_p(z)$ functions here. As an example, the dependences
$ \varphi_q(z)$ at  $d/l=1$ are shown in Fig.~\ref{fig3} (we use the
values of $C^p$, calculated below)

To find the normalization factor we rewrite the components
of the displacement vector in the form
\begin{eqnarray}
u_i({\bf r},z,t)=\sum_{\bf q} C^p f_i^p(z) e^{i {\bf q r}}
(b_{\bf q}+b_{-{\bf q}}^+)
\label{20}
\end{eqnarray}
and substitute the expression (\ref{20}) into the common expression
for the elastic energy. Comparing the result with Eq.
(\ref{13}), we obtain the following expression for the factors
$C^p$
\begin{equation}
C^p= (S v I^p)^{-1/2},
\label{21}
\end{equation}
where
\begin{equation}
I^p= 4 q \int_0^\infty d z \rho(z)  (|f_x^p(z)|^2 + |f_z^p(z)|^2).
\label{22}
\end{equation}
We note, that the $q$ dependence of $I^p$ ( and $C^p$ correspondingly)
is reduced to the dependence on the parameter $q a$.

Rewriting Eq.(\ref{19}) in terms of $b$ operators and
substituting the result into Eq.(\ref{15}),
we found the expression for the matrix elements in Eq. (\ref{14})
\begin{equation}
g_{q m} = i (4\pi \beta e /\epsilon \sqrt{v I^p}) \chi_p(z_m).
\label{23}
\end{equation}

The velocity renormalization $\Delta v$ and the absorption
coefficient  $\Gamma $ for the elastic wave coupled to the
two-dimensional electrons (the interaction is described by
Eq.(\ref{14})) are given by the equation
\begin{equation}
{\Delta v \over v} - i {\Gamma \over q}=
{1\over v q} g^*_{m q} D_{m m^{'}}(q,\omega_q)  g_{m^{'} q},
\label{24}
\end{equation}
where $D_{m m^{'}}$ is the electron density-density response
function. The values of $D$ in the random phase approximation
can be found from the equation
\begin{equation}
\hat{D}(q,\omega )= (\hat{I}- \hat{D}^{(0)}(q,\omega )
\hat{V}(q))^{-1}
\hat{D}^{(0)}(q,\omega ),
\label{25}
\end{equation}
where
\begin{equation}
V_{m, m^{'}}(q) = {2\pi e^2\over \epsilon q}(\delta_{m m^{'}} +
(1-\delta_{m m^{'}}) e^{- q d}) \ -
\label{26}
\end{equation}
the Fourier components of the Coulomb interaction,
$D^{(0)}$ is the density-density response function for the system
without the Coulomb interaction. The quantities
$D^{(0)}$ can be expressed through the longitudinal conductivity
of the electrons
\begin{equation}
D^{(0)}_{m m^{'}}(q,\omega )= -{i q^2\over \omega e^2}
\sigma^{m m^{'}}_{ x x} (q, \omega ),
\label{27}
\end{equation}
where $\sigma^{11}=\sigma^{22}$ are the diagonal with respect to the
electron layers,
$\sigma^{12}=\sigma^{21}$, nondiagonal with respect to the layers
components of the conductivity. Here and below we consider the case of
two equivalent electron layers. We should note, that the nondiagonal
components of conductivity are equal to zero for the electron gas in the
random phase approximation, while they may be nonzero for the composite
fermion gas due to the interlayer statistical interaction (this case
is considered in the next section)

Substituting Eqs. (\ref{23}, \ref{25}-\ref{27}) into Eq.
(\ref{24}), we obtain
\begin{eqnarray}
{\Delta v\over v }  - i {\Gamma \over q}=
\alpha_+
{-i \sigma_{xx}^+(q,vq)/\sigma_M^+ \over 1 + i \sigma_{xx}^+(q,v q)
/\sigma_M^+}  \cr +
\alpha_-
{-i \sigma_{xx}^-(q,vq)/\sigma_M^- \over 1 + i \sigma_{xx}^-(q,v q)
/\sigma_M^-} ,
\label{28}
\end{eqnarray}
where $\sigma_{xx}^\pm = \sigma_{xx}^{11} \pm \sigma_{xx}^{12}$,
$\sigma_M^\pm = v\epsilon/2 \pi  (1\pm \exp(-q d))$,
\begin{equation}
\alpha_\pm = {4\pi \beta ^2\over \epsilon v^2 I^p}
{|\chi_p(a)\pm\chi_p(-a)|^2 \over 1\pm \exp(- q d)}.
\label{29}
\end{equation}

The $\alpha_\pm$ functions play the role of the piezoelectric coupling
constants, which are introduced for the consideration of the interaction of
the SAW with the two-dimensional electrons. One can see from Eq. (\ref{29}),
that  $\alpha_+$ coefficient is nonzero for the type I mode only, while
the  $\alpha_-$ coefficient is nonzero for the type II mode only.

\begin{figure}
\narrowtext
\centerline{\epsfig{figure=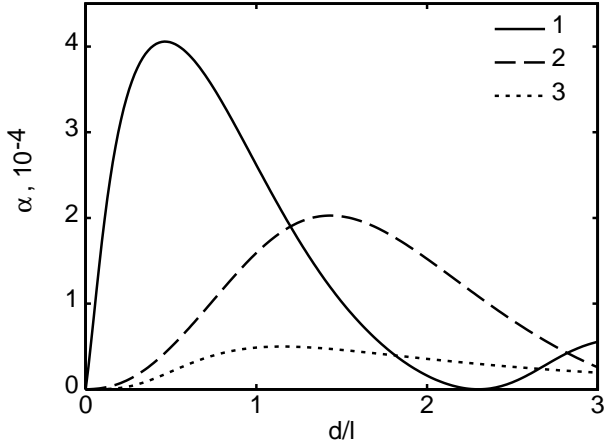,width=8cm}}
\vspace{0.5cm}
\caption{
Dependence of the  piezoelectric coupling constant on
the parameter $d/l$;
1, $\alpha_+$ for the type I mode;
2, $\alpha_-$ for the type II mode;
3, $\alpha_-$ for the transverse mode [5].}
\label{fig4}
\end{figure}

The dependences of $\alpha_+$ for the I mode and $\alpha_-$ for the
II mode versus the parameter $d/l$ are shown in Fig.~\ref{fig4}.
The parameters
$\beta=4.5 \cdot 10^{4}$ dyn$^{1/2}$/cm, $\epsilon $=12.5 are used.
The dependence of $\alpha_-$ on $d/l$ for the transverse mode considered
in Ref. \cite{5} is also shown in Fig.~\ref{fig4}.
One can see from the dependences presented, the interaction of the electrons
with the elastic modes, elliptically polarized in the sagittal plane is much
stronger then the interaction with the transverse mode.
Note, that the case considered in \cite{5} has the advantage, that
there is only one transverse nonuniform mode (if the thickness of the
central layer is not very large comparing to the wavelength), and its
frequency is lower than the frequencies of the bulk modes.
(For the case considered here the bulk transverse mode polarized
along the [1$\bar{1}$0] direction has the frequency, which is lower
then the frequencies of the I and II modes.

\section{Elastic mode velocity shift under the phase transition in
the fractional quantum Hall system}

Let us apply the results obtained in the previous section to the study
of a possibility of the observation of phase transitions in the
double-layer fractional quantum Hall systems.
To describe  the quantum Hall system
we use the
composite fermion approach
(the Chern-Simons fermionic model, developed in Ref.\cite{8}
for the double-layer system). Within such an approach the
fractional quantum Hall system is modelled as a gas of the composite
fermi-particles which carry the auxiliary statistical charge and the
flux of the statistical gauge field. For the double-layer
model two types of the statistical charges corresponding to two
layer and two types of the gauge fields are introduced.
In  general case the composite quasiparticles carry the fluxes of the
both types, namely the even number $\psi$ of the flux quanta of the
statistical field corresponding to their statistical charges and
the integer number $s$ of the flux quanta of the field, which corresponds
to the statistical charges in the opposite layer.
This model at  $s\ne 0$ corresponds the states described by the Halperin
wave function \cite{9}. We refer the phase with $s\ne 0$ as the
phase with the interlayer statistical interaction.

In the average field approximation the partial screening of the external
magnetic field $B$ emerges due to the influence of the statistical fields
\begin{equation}
B_{\rm eff}= B (1-\nu (\psi+s)),
\label{30}
\end{equation}
where $\nu$ is the filling factor calculated for one layer.
The fractional quantum Hall effect corresponds to the filling factors,
for which the value of $B_{\rm eff}$ corresponds to the integer number
$N$ of the filled Landau levels:
\begin{equation}
\nu = {N\over N(\psi+s)\pm 1},
\label{31}
\end{equation}
where the upper sign describes the case of
$B_{\rm eff}>0$, and the lower sign - the case of
$B_{\rm eff}<0$.
One can see from Eq. (\ref{31}), that certain fixed filling factors
may correspond to different sets of the $\psi$ and  $s$ parameters
(which describe different phases).

To analyze the incompressible states (which correspond to the
filling factors (\ref{31})) it is convenient to express the quantities
$\sigma_{xx}$ through the polarization tensor components
$\hat{\Pi}$:
\begin{equation}
\sigma_{xx}^{+(-)} = -{i\over \omega } \Pi^{+(-)}_{x x}=
-{i\omega \over q^2} \Pi^{+(-)}_{0 0} ,
\label{32}
\end{equation}
where $\Pi^{+(-)}=\Pi^{11}\pm \Pi^{12}$,
$\Pi^{11}$, $\Pi^{12}$ are the diagonal and nondiagonal with respect to
the layers components of the polarization tensor.
The values of $\Pi_{00}^{+(-)}$ are found to be
\begin{eqnarray}
\Pi_{00}^{+(-)}= -{e^2 q^2\over 2 \pi \omega_c} {S_0\over \Delta^{+(-)} },
\label{33}
\end{eqnarray}
where
\begin{eqnarray}
\label{34}
S_0=\Sigma_0 - {m^*-m_b\over m^* N} (\Sigma_0(\Sigma_2+N)-\Sigma_1^2),\\
\label{35}
\Delta^{+(-)}=(1-(\psi\pm s) \Sigma_1)^2 \cr -(\psi\pm s)^2 \Sigma_0
(\Sigma _2+N)-  {m^*-m_b\over m^* N} F,\\
\label{36}
F=\Sigma_2+N + (\omega/\omega_c)^2 S_0,
\end{eqnarray}
\begin{eqnarray}
\Sigma _j=({\rm sgn}(B_{\rm eff}))^j  \cr \times
e^{-x}\sum_{n=0}^{N-1} \sum_{m=N}^\infty {n!\over m!}
{x^{m-n-1}(m-n)\over {(\omega /\omega _c)^2 - (m-n)^2}}  \cr
\times [L_n^{m-n}(x)]^{2-j}
[ (m-n-x)L_n^{m-n}(x)  \cr + 2x{dL_n^{m-n}(x)\over dx} ] ^j.
\label{37}
\end{eqnarray}
In Eqs.(\ref{33}-\ref{37})
$\omega_c=2\pi n_0/m^* N$ is the effective cyclotron frequency,
$x=(q\lambda _{\rm eff} )^2/2$, where
$\lambda_{\rm eff}=(N/2\pi n_0)^{1/2}$ is the effective
magnetic length,
$L_n^{m-n}(x)$ is the generalized Laguerre polynomial,
$m^*$  is the effective mass of the composite fermions,
$m_b$ is the band mass of the electrons,
$n_0$ is the average electron density.
We used the modified random phase approximation \cite{10} for the
calculation
of $\Pi_{00}^{+(-)}$.

Substituting Eqs.(\ref{32},\ref{33}) into  Eq.(\ref{28}) we find
\begin{equation}
{\Delta v\over v} = \alpha_+ {E_q^+ S_0\over \Delta_+ - E_q^+ S_0}+
\alpha_- {E_q^- S_0\over \Delta_- - E_q^- S_0}
\label{38}
\end{equation}
where $E_q^{+(-)} = (e^2 q/ \epsilon \omega _c)(1\pm \exp(- q d))$
(the absorption coefficient $\Gamma$ is equal to zero for the
incompressible states).

Let us consider the filling factor $\nu =1/5$.
If there is no interlayer interaction, this filling factor
corresponds to the parameters
$\psi=4$, $s=0$, $N=1$.
When the interlayer distance becomes smaller a transition to the
phase $\psi=2$, $s=2$, $N=1$ may take place.
One can see from Eqs. (\ref{34}-\ref{37}),
the first term in Eq. (\ref{38}) is not changed under such a transition.
Therefore, the value of $\Delta v$ for the type I mode remains unchanged.
On the contrary the jump of the phase velocity takes place for the
mode of the type II (in this case the dependence on $\Delta_-$ which
is the function of $\psi - s$ survives in Eq. (\ref{38}))
The jump can be observed under the interlayer distance variation,
but it can be hardly realized in experiments. But the effect can be
observed indirectly if one measures the dependence of $\Delta v$ on
the wave vector. If one use the elastic mode
velocity at $\nu=1$ as the bare value (at $\nu=1$  the bare shift
is determined by the same Eq. (\ref{38}) with the parameters
$m^*=m_b$, $\psi =s=0$, $N=1$), the dependences of the relative shift
on the wave vector distinguish qualitatively for the cases
$s=0$ and $s\ne 0$.

\begin{figure}
\narrowtext
\centerline{\epsfig{figure=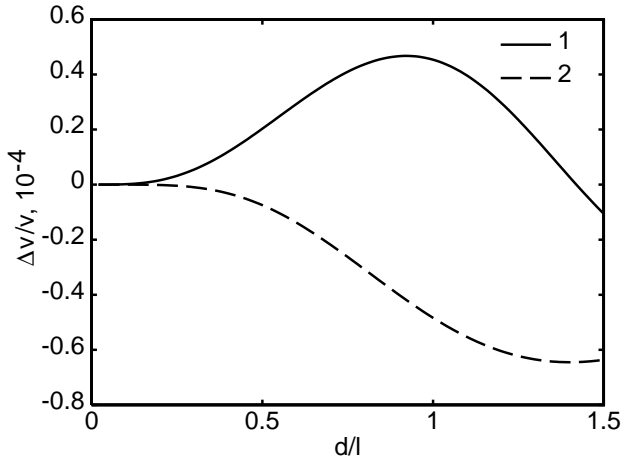,width=8cm}}
\vspace{0.5cm}
\caption{
Dependence of the  velocity shift on the inverse wavelength for the
type II mode at the filling factor $\nu=1/5$;
1, the phase $\psi =4$, $s=0$;
2, the phase  $\psi =2$, $s=2$.}
\label{fig5}
\end{figure}

Fig.~\ref{fig5} illustrates this behavior. The dependences $\Delta v/v$ on
the inverse wavelength
(where $\Delta v$ is the difference between the velocities at
$\nu=1/5$ and $\nu=1$) are shown.
We use the parameters
$n_0=10^{11}$ cm$^{-2}$,
$d=500 \AA$,
$m_b=0.07 m_e$, $m^*=4 m_b$ ($m_e$ is the electron mass) for the
calculations.

One can see from Eq.(\ref{31}), that
the phase transition at certain fixed filling factors
may be accompanied by the change of the sign of
$B_{\rm eff}$.
For instance, at $\nu =2/7$ the phase without the interlayer statistical
interaction corresponds to the parameters
$\psi=4$, $s=0$, $N=2$ ($B_{\rm eff}<0$), while the
phase with the interlayer statistical interaction may correspond to
the parameters  $\psi=2$, $s=1$, $N=2$ ($B_{\rm eff}>0$).
The transition between these phases results in
a jump of the velocity for the modes of the both types.
But the qualitative behavior of $\Delta v(q)$ remains
practically unchanged.

\begin{figure}
\narrowtext
\centerline{\epsfig{figure=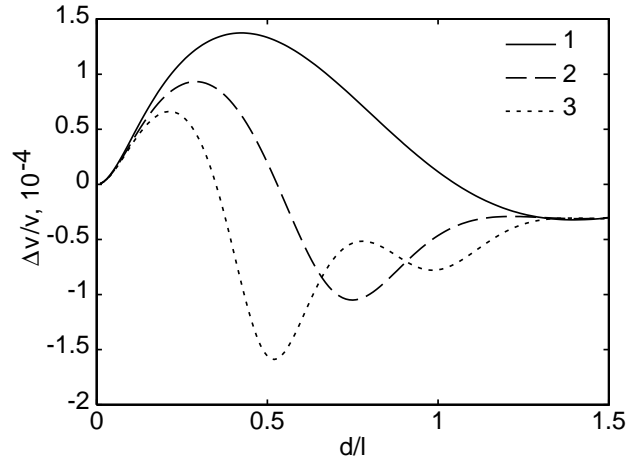,width=8cm}}
\vspace{0.5cm}
\caption{
Dependence of the  velocity shift on the inverse wavelength for the
type I mode;
1, $\nu =1/3$;
2, $\nu =2/5$;
3, $\nu =3/7$.}
\label{fig6}
\end{figure}

The mode of the type I,  for which the maximum of the coupling is
shifted to the long wave region, can  be used for the indirect
observation of the dependence of the effective magnetic length on
the filling factor.
The dependences  $\Delta v(d/l)$ (the value of $\Delta v$ is calculated
relative the velocity for the mode with the same $l$ at $\nu=1$) are shown
in Fig.~\ref{fig6} at $\nu =1/3$, 2/5, 3/7 ($\psi =2$ and $N=1,2,3$ correspondingly).
Here specify the case of $s=0$.
One can see from Fig.~\ref{fig6}, the value of $\Delta v$
oscillates as the function of  $1/l$.
The period of the oscillations is reduced when $\nu $ approaches to
the 1/2 value. It reflects that the effective magnetic length is
increased for this sequence of the filling factors.

Thus, the interaction of the double-layer electron system which is realized
at the interfaces of the wide quantum well in the
$Al Ga As - Ga As - Al Ga As$ heterostructure with nonuniform elastic
modes, localized in the central layer of the heterostructure and
elliptically polarized in the sagittal plane, is studied theoretically.
The dependence of the piezoelectric coupling constant on the ratio
between the thickness of the quantum well and the wavelength is found.
It is shown, that the coupling constant increases under the decreasing of
the wavelength and it reaches the maximal value at the wavelength which is
of order of thickness of the $Ga As$ layer (the concrete value of this
wavelength is determined by the type of the nonuniform mode).
The effect considered can be used for an experimental study of the
dynamical properties of two-dimensional electron layers at large
wave vectors, for which the interaction of the electrons with the
surface acoustic wave is suppressed exponentially due to the finite
value of the distance between the electron layer and the surface of the
sample.
The renormalization of the phase velocity of the nonuniform elastic
modes coupled to the double-layer fractional quantum Hall system
is calculated. It is shown, that the transition of the Hall system into the
state described by the Halperin wave function results in that the  dependence
of the velocity shift on the wave vector is modified quantitatively
for certain nonuniform elastic modes.

\end{multicols}
\end{document}